\documentclass[12pt]{article}

\usepackage{bm}
\usepackage{amsmath,amssymb,amsthm}
\usepackage{mathbbol,bbm}

\newcommand{\bi}[1]{\boldsymbol{#1}}
\newcommand{\idty}{\Eins}
\DeclareMathOperator{\Tr}{Tr}
\newcommand{\s}[1]{\mathsf{#1}}
\newcommand{\C}{\Bbb{C}}
\newcommand{\Z}{\Bbb{Z}}
\newcommand{\R}{\Bbb{R}}
\newcommand{\T}{\Bbb{T}}
\newcommand{\alg}{\mathcal{A}}
\newcommand{\gicar}{\alg^{\mathrm{gi}}}
\newcommand{\hil}{\mathcal{H}}
\newcommand{\matr}{\mathcal{M}}
\newcommand{\oppa}{\mathcal{X}}
\newcommand{\opppa}{\mathcal{Y}}
\newcommand{\refdens}[1]{\rho^{(#1)}_{\mathcal{X},\Theta}}
\newcommand{\trefdens}[1]{\tilde\rho^{(#1)}_{\mathcal{X},\Theta}}
\newcommand{\refpart}[1]{\mathcal{X}^{(#1)}_{\Theta}}
\newcommand{\tpl}[1]{\boldsymbol{#1}}
\newcommand{\eset}{\mathcal{E}}
\providecommand{\abs}[1]{\lvert#1\rvert}

\begin{document}

\begin{center}
\textbf{\large
 Robustness of dynamical entropy} \\[6pt]
{\large
 M.~Fannes\footnote{E-mail: mark.fannes@fys.kuleuven.ac.be},
 B.~Haegeman\footnote{E-mail: bart.haegeman@fys.kuleuven.ac.be, Research Assistant
 FWO-Vlaanderen}}
 and
{\large
 D.~Vanpeteghem\footnote{E-mail: dimitri.vanpeteghem@fys.kuleuven.ac.be, Research
 Assistant FWO-Vlaanderen}} \\[6pt]
\emph{Instituut voor Theoretische Fysica, K.U.Leuven,
Celestijnenlaan~200D, B-3001 Heverlee, Belgium}
\end{center}
\bigskip

\noindent \textbf{Abstract:}
When quantifying the mixing properties of a quantum dynamical system in terms of dynamical entropy, the following scheme appears natural: observe the state of the system at regular time intervals while it evolves and determine the entropy produced over time. It is clear that this entropy will not only depend on the type of dynamics, but also on the type of observations. Intuitively, one can expect that some measurements are better suited than others to reveal information about the dynamics, whereas many will generate undesirable noise. In this paper, we show for two widely used model systems that the dynamical entropy is rather robust in this respect. More precisely, general local positive operator-valued measurements may be restricted to von~Neumann type measurements for the shift on a quantum spin chain and gauge-invariant ones for the shift on a Fermion chain.
\medskip

\section{Introduction}

Classical dynamical systems that admit a generating partition are, up to isomorphism, classified by the value of their Kolmogorov--Sinai invariant, see~\cite{P}. This is achieved by mapping the dynamical system on a shift dynamics on symbolic sequences written in an alphabet with sufficiently many letters. An isomorphism is obtained because the encoding essentially maps a phase space point into a sequence. One cannot hope to extend this procedure to quantum dynamical systems as there is no underlying phase space. Already the shifts on quantum spin chains --- the putative standard models for quantum dynamical systems in discrete time --- show this difficulty: local observables commute as soon as they are sufficiently pulled apart so that their domains of dependency become disjoint, while generally observables at largely separated times never commute. Also two shifts on quantum spin chains with the same mean entropy don't have to be isomorphic, even the type of the associated von~Neumann algebra can be different.

It is therefore not surprising that the Kolmogorov--Sinai invariant extends in various distinct ways to quantum dynamical systems, and different extensions feel different properties of the system. The best-known extension is the CNT~dynamical entropy (Connes--Narnhofer--Thirring), based on decompositions of the reference state, see~\cite{CNT}. This construction can be rephrased in terms of a coupling with a classical dynamical system, see~\cite{ST}. A second approach, the ALF~entropy (Alicki--Lindblad--Fannes), is based on operational partitions of unity, see~\cite{AF1,AF2}. It arises by alternating generalised measurements with the dynamics. The CNT~entropy seems to encode rather the commutative aspects of the dynamics while the ALF~construction is more sensitive to non-commutativity.

Model systems are known for which these entropies yield very different results. Even such extremes as zero for CNT and infinity for ALF occur, viz.\ for free shifts, for Powers--Price shifts~\cite{AN} and for classical stochastic systems~\cite{FH}. Even systems with a clearer physical input, such as shifts on spin chains and free evolutions on CAR~algebras (canonical anti-commutation relations) produce different CNT and ALF entropies. For a shift on a chain with $d$-dimensional single site space, the expected classical value, the entropy density $\sigma(\omega)$, is returned by CNT but in the case of ALF an extra term $\ln d$ shows up.

We shall in this paper investigate robustness properties of the ALF~entropy. As mentioned above, the mathematical construction involves besides the dynamics a generalised measurement. However, there is a lot of freedom in choosing the corresponding partitions of unity and, in principle, repeated measurements can in themselves generate entropy. It is, in general, an open problem to decide what the impact is on the ALF~entropy.

We shall restrict here our attention to two basic models: the shifts on a spin on a Fermion chain. For the spin chain, we shall show that instead of using generalised measurements as described by general partitions of unity and their corresponding POVM's (positive operator-valued measures) it suffices to consider von~Neumann type measurements, i.e.\ projection-valued partitions. The idea of modelling a quantum dynamical systems by its multi-time correlation functions associated with a projection-valued measurement goes at least back to proposals by Feynman and Gell-Mann. For the shift on the CAR~chain, we shall show that we may restrict our attention to partitions in gauge-invariant elements. Such elements correspond to second quantised observables.

We conclude this introduction with a lemma that will prove useful in obtaining upper bounds for the entropy. Recall that a size-$k$ operational partition of unity on a Hilbert space $\hil$ is a collection $\oppa = \{x_i \mid i=1,2,\ldots,k\}$ of operators on $\hil$ such that
\begin{equation}
  \sum_{i=1}^k x_i^*x_i = \idty.
\label{eq:partunit}
\end{equation}
This definition straightforwardly generalises to a unital algebra of operators. Let, moreover, $\omega$ be a density matrix on $\hil$. We can then introduce the $k$-dimensional correlation matrix $\rho_\oppa$ with $(i,j)$-th entry
\begin{equation*}
 \rho_\oppa(i,j) := \Tr (\omega\, x_j^*x_i ),\quad i,j=1,2,\ldots k.
\end{equation*}
Obviously, $\rho_{\oppa}$ is a density matrix. We finally need the  von~Neumann entropy of a density matrix $\sigma$
\begin{equation*}
 \s S(\sigma) := - \Tr \sigma\ln\sigma.
\end{equation*}

\textbf{Lemma} \ Let $\omega$ be a density matrix on a finite dimensional Hilbert space $\hil$ and let $\oppa$ be a size-$k$ partition of unity on $\hil$. Then
\begin{equation*}
 \s S(\rho_\oppa) \le \s S(\omega) + \ln\dim(\hil).
\end{equation*}
\smallskip

Denote the spectral decomposition of $\omega$ by $\omega = \sum_i \lambda_i\, |\phi_i\rangle \langle\phi_i|$ and consider the coupled system $\hil\otimes\C^k$. The vectors $\eta_i := \sum_{j=1}^k x_j\phi_i \otimes e_j$ form an orthonormal set so that the state $\sum_i \lambda_i\, |\eta_i\rangle \langle\eta_i|$ on the extended system has the same entropy as $\omega$. The partial trace of this density matrix on $\C^k$ equals $\rho_\oppa$ while on $\hil$ it has rank at most $\dim(\hil)$. Applying the triangle inequality for the entropy finishes the proof.

\section{Dynamical entropy}
\label{sec:dynentr}

In this section, we briefly recall the construction of the quantum dynamical entropy as defined in~\cite{AF1}. More details can be found in~\cite{AF2}. A quantum dynamical system is given by a triple $(\alg,\Theta,\omega)$. $\alg$ is called the algebra of observables, the automorphism $\Theta: \alg\to\alg$ is the single step dynamical map and the state $\omega$ on $\alg$ is the reference state, invariant under $\Theta$.

As in the classical Kolmogorov-Sinai construction an initial partition $\oppa$ gets refined under the dynamics. For two partitions $\oppa = \{x_i \mid i=1,2,\ldots,k\}$ and $\opppa = \{y_j \mid j=1,2,\ldots,\ell\}$, we define the ordered composition of $\oppa$ and $\opppa$ as $\oppa\lor\opppa := \{x_iy_j \mid i=1,2,\ldots,k,\ j=1,2,\ldots,\ell\}$. The evolution $\Theta$ of a partition $\oppa$ is $\Theta(\oppa) := \{\Theta(x_i) \mid i=1,2,\ldots,k\}$. This gives us the $N$-steps refinement of partition,
\begin{equation}
\label{eq:refipart}
 \refpart{N} := \Theta^{N-1}(\oppa) \lor\ldots\lor \Theta(\oppa) \lor \oppa
\end{equation}
with $\tpl{j}$-th element
\begin{equation}
\label{eq:elempart}
 \refpart{N}(\tpl{j}) = \Theta^{N-1}(x_{j_{N-1}}) \cdots
\Theta(x_{j_1})x_{j_0},\qquad \tpl{j} = (j_0,j_1,\ldots,j_{N-1}).
\end{equation}
With a refined partition $\refpart{N}$ and a state $\omega$ we now can associate the $N$-steps correlation matrix $\refdens{N}$. Explicitly,
\begin{equation}
\label{eq:refidens}
 \refdens{N}(\tpl{i},\tpl{j}) :=
 \omega(\refpart{N}(\tpl{j})^*\refpart{N}(\tpl{i})).
\end{equation}

The von~Neumann entropy $\s S$ of these refined density matrices is used to construct the entropy of a partition
\begin{equation*}
 \s h(\Theta,\omega,\oppa) :=  \limsup_{N\to\infty}
 \frac{1}{N}\, \s S(\refdens{N}).
\end{equation*}
and the quantum dynamical entropy of the system $(\alg,\Theta,\omega)$,
\begin{equation}
\label{eq:dynentr}
 \s h(\Theta,\omega) :=  \sup_\oppa \s h(\Theta,\omega,\oppa)
\end{equation}
The supremum over the possible partitions of unity $\oppa$ deserves special attention. These partitions correspond to the measurements allowed to extract information from the dynamics. The dynamical entropy therefore not only depends on the dynamical system $(\alg,\Theta,\omega)$, but also on the class of allowed partitions.

\section{Shift on a spin chain}

As a first example, we consider the shift on a quantum spin chain. In~\cite{AF1,AF2}, the entropy~(\ref{eq:dynentr}) was computed allowing arbitrary partitions in local elements. We prove that we obtain the same result if we restrict our attention to partitions in orthogonal projections, corresponding to the standard von~Neumann type measurements.

The observables of a single spin form the algebra $\matr_d$  of $d\times d$ matrices. The spins in a finite subvolume $\Lambda$ of $\Z$ are then described by $\alg_\Lambda := \bigotimes_\Lambda \matr_d$. The natural embedding of $\alg_{\Lambda_1}$ in $\alg_{\Lambda_2}$ for $\Lambda_1 \subset \Lambda_2$ obtained by tensoring elements of $\alg_{\Lambda_1}$ with the identity of $\alg_{\Lambda_2\setminus\Lambda_1}$ allows us to construct the algebra $\alg$ of quasi-local observables of the quantum spin chain
\begin{equation*}
 \alg := \bigotimes_\Z \matr_d = \overline{\bigcup_{\Lambda\subset\Z}
 \alg_\Lambda},
\end{equation*}
where the bar denotes the norm closure. For $n\in\Z$, denote by $\imath_n$ the canonical injection of $\matr_d$ in the $n$-th factor of $\alg$. The dynamical map $\Theta$ is the shift automorphism on $\alg$, defined by $\Theta(\imath_n(A)) = \imath_{n+1}(A)$ with $A\in\matr_d$. The reference state $\omega$ is an arbitrary translation invariant state on $\alg$, meaning that $\omega\circ\Theta = \omega$.

\noindent
\textbf{\large Entropy of the shift}
\smallskip

Consider now a partition $\oppa$ in local elements. Because of shift-invariance we can assume that they live on the interval $[1,M]$. The dynamics shifts these elements to the right so that after $N$ time steps, the refined partition lives on $[1,M+N]$. The algebra $\alg_{[1,M+N]}$ is just the algebra of $d^{M+N}$ dimensional matrices. Therefore, using the lemma
\begin{equation*}
 \s S(\refdens{N}) \le \s S(\omega_{[1,M+N]}) + \ln d^{M+N}.
\end{equation*}
Here, $\omega_{[1,M+N]}$ denotes the density matrix on $\C^{d^{M+N}}$ that defines the restriction of the reference state $\omega$ to $\alg_{[1,M+N]}$ realised as the algebra of matrices of dimension $d^{M+N}$. Dividing both sides by $N$ and taking the limit $N\to\infty$, we obtain
\begin{equation}
\label{ub}
 \s h(\Theta,\omega,\oppa) \le \sigma(\omega) + \ln d.
\end{equation}
We shall show that this inequality is saturated when we start out with a suitable partition in orthogonal projections.

Feynman and Gell-Mann proposed to encode the dynamics of a quantum system in terms of time-ordered multi-time correlation functions associated to a von~Neumann measurement. Let $\{P_i \mid i=1,2,\ldots,k\}$ be a decomposition of the identity in orthogonal projections
\begin{equation*}
 P_i = P_i^* = P_i^2
 \qquad\text{and}\qquad
 \sum_{i=1}^{k} P_i = \idty.
\end{equation*}
Denoting by $P_i(t)$ the evolution of $P_i$ during a time $t$, i.e.\ $P_i(t) = U^*(t) P_i U(t)$ where $\{U(t) \mid t\in\R\}$ is the unitary time evolution on the Hilbert space $\hil$ in the case of standard quantum mechanics, these correlation functions are
\begin{equation*}
 (\tpl{i},\,\tpl{t}) \mapsto \langle \cdots P_{i_1}(t_1) P_{i_0}(t_0)
 P_{i_1}(t_1) \cdots\rangle,
\end{equation*}
with $\tpl{i} = (i_0,\ldots,i_{N-1})$ and $\tpl{t} = (t_0,\ldots,t_{N-1}),\ t_{i_n}<t_{i_{n+1}}$. If the $P_i$ are one-dimensional, then the dynamics can be reconstructed using Wigner's Theorem.

Choose an orthonormal basis $\{|e_i\rangle\}$ for $\C^d$ and its associated Fourier basis
\begin{equation*}
 |f_j\rangle := \frac{1}{\sqrt{d}} \sum_{k=1}^d \exp \bigl(2\pi ijk/d\bigr)
 |e_k\rangle.
\end{equation*}
Next, consider the projectors $p_i := |e_i\rangle \langle e_i|$ and $q_j := |f_j\rangle \langle f_j|$. Both sets $\{p_i\}$ and $\{q_j\}$ are decompositions of the identity such that $\oppa := \{p_i\otimes q_j\,|\,i,j=1,\ldots d\}$ is a partition of unity. We now compute the corresponding refined correlation matrices
\begin{align*}
 &\refdens{N}(\tpl{i}\tpl{k},\tpl{j}\tpl{\ell}) \\
 &\qquad= \omega\bigl( p_{i_0}\!\otimes\!q_{k_0}\,
        \Theta(p_{i_1}\!\otimes\!q_{k_1}) \cdots
        \Theta^{N-1}(p_{i_{N-1}}\!\otimes\!q_{k_{N-1}})\,
        \Theta^{N-1}(p_{j_{N-1}}\!\otimes\!q_{\ell_{N-1}})\, \cdots  \\
 &\phantom{\qquad= \omega\bigl(} \Theta(p_{j_1}\!\otimes\!q_{\ell_1}) \,
        p_{j_0}\!\otimes\!q_{\ell_0} \bigr) \\
 &\qquad= \omega\bigl( p_{i_0}p_{j_0} \otimes
        q_{k_0}p_{i_1}p_{j_1}q_{\ell_0} \otimes\cdots\otimes
        q_{k_{N-2}}p_{i_{N-1}}p_{j_{N-1}}q_{\ell_{N-2}} \otimes
        q_{k_{N-1}}q_{\ell_{N-1}} \bigr) \\
 &\qquad= \delta_{\tpl{i}\tpl{j}}\, \delta_{k_{N-1}\ell_{N-1}} \,
     \omega( p_{i_0} \otimes q_{k_0}p_{i_1}q_{l_0}
     \otimes\cdots \otimes
     q_{k_{N-2}}p_{i_{N-1}}q_{\ell_{N-2}} \otimes q_{k_{N-1}} ) \\
 &\qquad= \delta_{\tpl{i}\tpl{j}} \delta_{k_{N-1}\ell_{N-1}} \,
          \omega( p_{i_0} \otimes |f_{k_0}\rangle\langle f_{\ell_0}|
          \otimes\cdots \otimes
         |f_{k_{N-2}}\rangle\langle f_{\ell_{N-2}}| \otimes q_{k_{N-1}} ) \\
 &\phantom{\qquad= \delta_{\tpl{i}\tpl{j}}}
             \frac{1}{d^{N-1}} \prod_{n=0}^{N-2}
             \exp\bigl( 2\pi i i_n(\ell_n-k_n)/d\bigr).
\end{align*}
{From} this $d^{2N}$-dimensional density matrix we split off the first $d$ dimensions (indexed by $i_0,j_0$) and the last $d$ dimensions (indexed by $k_{N-1},\ell_{N-1}$). Denote the density matrices reduced to these $d^{2(N-1)}$ dimensions by $\trefdens{N}$ and to the remaining $d^2$ dimensions by $\rho^r$. By the triangle inequality,
\begin{equation*}
 \s S(\refdens{N}) \ge \s S(\trefdens{N}) - \s S(\rho^r)
 \ge \s S(\trefdens{N}) - 2\ln d.
\end{equation*}
We now compute the components of the density matrix $\trefdens{N}$, using the notations $\tilde{\tpl{i}} = (i_1,i_2,\ldots,i_{N-1})$ and $\tilde{\tpl{k}} = (k_0,k_1,\ldots,k_{N-2})$, and similarly $\tilde{\tpl{k}}$ and $\tilde{\tpl{\ell}}$.
\begin{align*}
 &\trefdens{N}(\tilde{\tpl{i}} \tilde{\tpl{k}}, \tilde{\tpl{j}}
 \tilde{\tpl{\ell}}) \\
 &\qquad= \frac{1}{d^{N-1}} \prod_{n=0}^{N-2}
   \exp\bigl( 2\pi i i_n(\ell_n-k_n)/d \bigr)\
  \delta_{\tilde{\tpl{i}} \tilde{\tpl{j}}} \,
 \omega\bigl( |f_{k_0}\rangle\langle f_{\ell_0}| \otimes \cdots
        \otimes |f_{k_{N-2}}\rangle\langle f_{\ell_{N-2}}| \bigr) \\
 &\qquad= \frac{1}{d^{N-1}} \delta_{\tilde{\tpl{i}} \tilde{\tpl{j}}} \,
 \omega\bigl(
   \big| \exp( -2\pi i\,\tilde{\tpl{i}}\cdot\tilde{\tpl{k}}/d )
   f_{\tilde{\tpl{k}}} \,\big\rangle
   \big\langle \exp( -2\pi i\,\tilde{\tpl{i}}\cdot\tilde{\tpl{\ell}}/d )
   f_{\tilde{\tpl{\ell}}} \,\big| \bigr).
\end{align*}
This matrix is diagonal in the indices $\tilde{\tpl{i}}$ and $\tilde{\tpl{j}}$, whereas for every $\tilde{\tpl{i}}$, the set
\begin{equation*}
\Bigl\{ \big| \exp( -2\pi i\,\tilde{\tpl{i}}\cdot\tilde{\tpl{k}}/d )
    f_{\tilde{\tpl{k}}}\,\big\rangle \,\Bigr|\, \tilde{\tpl{k}}\in
 \{1,2,\ldots, d\}^{N-1} \Bigr\}
\end{equation*}
is an orthonormal basis for $\C^{d(N-1)}$. Therefore, its entropy equals
\begin{equation*}
 \s S(\trefdens{N}) = (N-1)\ln d + \s S(D_{\omega}^{(N-1)}),
\end{equation*}
where $D_{\omega}^{(N)}$ is the density matrix of $\omega$ reduced to an interval of $N$ sites. Finally, dividing by $N$ and taking the limit $N\to\infty$,
\begin{equation*}
 \s h(\Theta,\omega,\oppa) \ge \lim_{N\to\infty} \frac{1}{N}
 \s S(\refdens{N}) = \lim_{N\to\infty} \frac{1}{N} \s S(\trefdens{N})
 = \ln d + \sigma(\omega).
\end{equation*}
We thus attain the upper bound in~(\ref{ub}).

\section{Shift on a Fermion chain}

As a second example we study the shift on the Fermion chain. The algebra of observables $\alg$ is now the algebra of canonical anticommutation relations (CAR). It is the C*-algebra generated by the identity and elements $\{a_k \mid k\in\Z\}$ satisfying the relations
\begin{equation*}
 a_ka_l + a_la_k = 0
 \qquad\text{and}\qquad
 a_k^*a_l + a_la_k^* = \delta_{k,l} \idty.
\end{equation*}
The dynamical map $\Theta$ is the shift automorphism given by $\Theta(a_k) = a_{k+1}$ and the reference state $\omega$ is the tracial state. It is uniquely determined by the condition $\omega(AB)=\omega(BA)$ for $A,B\in\alg$ and explicitly given on ordered monomials by
\begin{equation*}
 \omega(a^*_{k_1} \cdots a^*_{k_n} a_{\ell_n} \cdots a_{\ell_1}) =
 2^{-n},\qquad k_1<\cdots<k_n,\ \ell_1<\cdots<\ell_n.
\end{equation*}
All other monomials have zero expectation.

To specify our restricted class of allowed partitions, we need the gauge-invariant subalgebra $\gicar$. For a scalar $\lambda\in\T := \{z \mid \abs{z}=1\}$ define the so-called gauge-automorphism $\alpha_\lambda$ by $\alpha_\lambda(a_k) := \overline\lambda a_k$. The GICAR algebra $\gicar$ is the subalgebra of $\alg$ invariant under all the gauge-automorphisms,
\begin{equation*}
 \gicar := \{ A \in \alg \mid \alpha_\lambda(A) = A
 \text{ for all }  \lambda \in \T \}.
\end{equation*}
An element $A := a_{k_1}^*\ldots a_{k_n}^*a_{\ell_m}\cdots a_{\ell_1}$, is mapped into $\alpha_\lambda(A) = \lambda^{n-m} A$. It belongs to $\gicar$ if and only if $m=n$. In fact, $\gicar$ is spanned by such elements.

The algebra $\alg$ is the abstract version of the algebra generated by e.g.\  Fermionic Fock creation and annihilation operators. The one-particle space is $\ell^2(\Z)$ and $a_k = a(e_k)$ where $e_k$ is the standard $k$-th basisvector in $\ell^2(\Z)$. The element $a_k$ is therefore the annihilation operator of the $k$-th mode in a chain of Fermionic modes. The elements of $\gicar$ are linear combinations of monomials that contain as many creation as annihilation operators. Such elements conserve the number of particles. If we realise our abstract algebra on the Fermionic Fock space, then the elements of $\gicar$ are second quantised local observables. The tracial state $\omega$ may be thought of as the most random state on $\alg$, namely the infinite temperature state.

Using the Jordan-Wigner isomorphism we can map the CAR algebra onto a quantum spin chain. For our application it will suffice to consider a subalgebra of the CAR algebra generated by $\{a_k \mid k\ge 1\}$ and map this algebra on the one-sided chain $\bigotimes_{k\ge 1}\matr_2$. This isomorphism is constructed as follows.

For $n\ge 1$, define
\begin{equation*}
 V_n := \prod_{k=1}^{n-1}(2a_k^*a_k-\idty),
\end{equation*}
and
\begin{equation*}
 E_{21}^{(n)} := V_na_n,\quad
 E_{12}^{(n)} := V_na_n^*,\quad
 E_{11}^{(n)} := a_n^*a_n\quad\text{and }
 E_{22}^{(n)} := a_na_n^*.
\end{equation*}
The sub-algebra $\alg_n$ generated by $\{a_k \mid 1\le k\le n\}$ is isomorphic to $\bigotimes_{k=1}^n \matr_2$ with matrix units
\begin{equation}
\label{eq:matrunit}
 E_{\bi\phi \bi\psi}^{[1,n]} := \prod_{k=1}^n E_{\phi_k\psi_k}^{(k)},
\end{equation}
where $\bi\phi = (\phi_1,\ldots,\phi_n)\in\{1,2\}^n$ and similarly for $\bi\psi$. An appropriate limit of this construction for $n\to\infty$ leads to the quantum spin chain $\bigotimes_{k\ge 1} \matr_2$.

The action of the gauge-automorphism $\alpha_\lambda$ on the matrix units $E_{\bi\phi \bi\psi}^{[1,n]}$ is
\begin{equation*}
 \alpha_\lambda \bigl( E_{\bi\phi \bi\psi}^{[1,n]} \bigr) =
 \lambda^{\Sigma\bi\psi-\Sigma\bi\phi}\, E_{\bi\phi \bi\psi}^{[1,n]},
\end{equation*}
where $\Sigma\bi\phi := \sum_{k=1}^n \phi_k$ and $\Sigma\bi\psi := \sum_{k=1}^n \psi_k$. The invariant elements are those for which $\Sigma\bi\phi = \Sigma\bi\psi$.

Sums like $\Sigma\bi\phi$ can take values $n,n+1,\ldots, 2n$. For each integer $0\le s\le n$,
\begin{equation*}
 \mathcal{F}_s^n := \mathrm{span}
 \Bigl\{ E_{\bi\phi \bi\psi}^{[1,n]} \,\Bigm|\,
 \Sigma\bi\phi = \Sigma\bi\psi = n+s \Bigr\}
\end{equation*}
is a $\ast$-algebra. Indeed, we have $\bigl( E_{\bi\phi \bi\psi}^{[1,n]} \bigr)^*  = E_{\bi\psi \bi\phi}^{[1,n]}$ and $E_{\bi\phi \bi\psi}^{[1,n]} E_{\bi\phi'\bi\psi'}^{[1,n]} = \delta_{\bi\phi' \bi\psi} E_{\bi\phi \bi\psi'}^{[1,n]}$ and thus $\Sigma\bi\phi = \Sigma\bi\psi' = n+s$. Moreover, the sum $\Sigma\bi\phi$ equals $n+s$ for $\binom{n}{s}$ elements $\bi\phi\in\{1,2\}^n$. Therefore, $\mathcal{F}_s^n = \matr_{\binom{n}{s}}$. One also has that $\mathcal{F}_s^n \mathcal{F}_t^n = 0$ for $s\neq t$. This leads to the direct sum decomposition,
\begin{equation*}
 \gicar_n := \gicar\cap\alg_n = \bigoplus_{s=0}^n \matr_{\binom{n}{s}}.
\end{equation*}

Finally, note that the effect of the shift $\Theta$ on an element~(\ref{eq:matrunit}) is not completely trivial. Using the notation
\begin{equation*}
 E_{\bi\phi \bi\psi}^{[m,n]} := \prod_{k=m}^n E_{\phi_k\psi_k}^{(k)},
\end{equation*}
where $\bi\phi,\bi\psi\in\{1,2\}^{n-m+1}$, one has
\begin{equation*}
 \Theta(E_{\bi\phi \bi\psi}^{[1,n]}) =
 (2a_1^*a_1-\idty)^{\Sigma\bi\phi-\Sigma\bi\psi}
 E_{\bi\phi \bi\psi}^{[2,n+1]}.
\end{equation*}
For gauge-invariant elements $E_{\bi\phi \bi\psi}^{[1,n]}$, $\Sigma\bi\phi = \Sigma\bi\psi$ and we simply get $\Theta(E_{\bi\phi \bi\psi}^{[1,n]}) = E_{\bi\phi \bi\psi}^{[2,n+1]}$.

\noindent
\textbf{\large Entropy of the shift}
\smallskip

An upper bound for the dynamical entropy can readily be obtained using the lemma at the end of the introduction. Fixing any local partition $\oppa$, i.e.\ a partition whose elements belong to a local algebra $\alg_n$ for $n$ large enough, and recalling that the  reference state is the tracial state, we obtain
\begin{equation}
\label{ub2}
 \s h(\Theta,\omega,\oppa) \le 2\ln2.
\end{equation}

In the following, we will construct a gauge-invariant partition of
unity which effectively realizes this upperbound. This partition
will have some nice properties which will allow us to map the
system on a Markov process. This process itself will be simplified
even further by a coarse graining of the state space. In the end,
the entropy for this process will be calculated and by
construction this will be the dynamical entropy for the particular
choice of gauge-invariant partition we made.

Consider the set
\begin{equation*}
 \oppa := \bigl\{ c_{\bi\phi \bi\psi} E_{\bi\phi \bi\psi}^{[1,M]}
            \,\bigm|\, (\bi\phi,\bi\psi)\in I_0 \bigr\},
\end{equation*}
where $E_{\bi\phi \bi\psi}^{[1,M]}$ are matrix units in the CAR~algebra, see~(\ref{eq:matrunit}), and where $c_{\bi\phi \bi\psi}$ are complex numbers to be determined later. The index set $I_0\subset\{1,2\}^{2M}$ can be chosen arbitrarily respecting the following two conditions. First, we want only gauge-invariant elements in the partition, meaning that $\Sigma\bi\phi = \Sigma\bi\psi$. Next, we impose that for every $(\phi_1;\psi_1,\ldots\psi_M )$ there exists at most one $(\phi_2,\ldots,\phi_M)$ for which $(\bi\phi,\bi\psi)\in I_0$. In other words, the index set $I_0$ is specified by a map $(\phi_1;\psi_1,\ldots\psi_M) \mapsto (\phi_2,\ldots,\phi_M)$ respecting gauge-invariance.

For $\oppa$ to be a partition of unity, see~(\ref{eq:partunit}), we have to ensure that
\begin{equation*}
 \idty = \sum_{(\bi\phi,\bi\psi)\in I_0} \abs{c_{\bi\phi \bi\psi}}^2\,
 \bigl( E_{\bi\phi \bi\psi}^{[1,M]} \bigr)^* E_{\bi\phi \bi\psi}^{[1,M]}
 = \sum_{(\bi\phi,\bi\psi)\in I_0} \abs{c_{\bi\phi \bi\psi}}^2\,
 E_{\bi\psi \bi\psi}^{[1,M]}
\end{equation*}
and thus that,
\begin{equation}
\label{eq:punitE}
 \sum_{\bi\phi\,:\,(\bi\phi,\bi\psi)\in I_0} \abs{c_{\bi\phi \bi\psi}}^2 = 1
 \quad \text{for all } \bi\psi\in\{1,2\}^M.
\end{equation}

The partition after $N$ refinements will still consist of elements proportional to matrix units, now living on $N+M$ sites. Such an element has the following structure
\begin{align}
 E_{(\phi_1,\ldots,\phi_{M+N}) (\psi_1,\ldots,\psi_{M+N})}^{[1,M+N]}
 &= E_{(\phi_{N+1},\ldots,\phi_{M+N}) (\bi t^{(N)},\psi_{M+N})}^{[N+1,M+N]}
    E_{(\phi_N,\bi t^{(N)}) (\bi t^{(N-1)},\psi_{M+N-1})}^{[N,M+N-1]} \cdots
\nonumber \\
 &\phantom{=\ }
    E_{(\phi_2,\bi t^{(2)}) (\bi t^{(1)},\psi_{M+1})}^{[2,M+1]}
    E_{(\phi_1,\bi t^{(1)}) (\psi_1,\ldots,\psi_M)}^{[1,M]},
\label{eq:refiexpl}
\end{align}
where $\bi t^{(n)}\in\{1,2\}^{M-1}$ for $n=1,\ldots,N$ and where we have used the notation $(\phi,\bi t)$ to indicate the concatenation of $(\phi)$ and $\bi t$. Due to the structure of the index set $I_0$, there is only one combination of elements of the initial partition $\oppa$ leading to a given element of the refined partition $\refpart{N}$. We will denote the latter by
\begin{equation*}
 \refpart{N} = \bigl\{ c_{\bi\phi \bi\psi} E_{\bi\phi \bi\psi}^{[1,M+N]}
                 \,\bigm|\, (\bi\phi,\bi\psi)\in I_N \bigr\}.
\end{equation*}
with the new index set $I_N\subset\{1,2\}^{2(M+N)}$.

The corresponding correlation matrix is
\begin{equation*}
 \refdens{N}(\bi\phi\bi\psi,\bi\phi'\bi\psi') =
 \frac{1}{2^{N+M}}
 \Tr \Bigl( \bigl( c_{\bi\phi\bi\psi} E_{\bi\phi\bi\psi}^{[1,M+N]} \bigr)^*
 c_{\bi\phi'\bi\psi'} E_{\bi\phi'\bi\psi'}^{[1,M+N]} \Bigr) =
 \frac{1}{2^{N+M}} \abs{c_{\bi\phi\bi\psi}}^2
 \delta_{\bi\phi\bi\phi'} \delta_{\bi\psi\bi\psi'},
\end{equation*}
and, as a given refinement can only be obtained in a single way, this correlation matrix is diagonal.

The problem has therefore been reduced to a dynamical entropy computation of a classical dynamical system. The probabilities on the diagonal can be written as in~(\ref{eq:refiexpl})
\begin{align}
 \refdens{N}(\bi\phi\bi\psi,\bi\phi\bi\psi)
 &= \frac{1}{2^{M+N}}
     \bigl| c_{(\phi_1,\bi t^{(1)}) (\psi_1,\ldots,\psi_M)} \bigr|^2
     \bigl| c_{(\phi_2,\bi t^{(2)}) (\bi t^{(1)},\psi_{M+1})} \bigr|^2
     \cdots
\nonumber \\
 &\phantom{\frac{1}{2^{M+N}}}
     \bigl| c_{(\phi_N,\bi t^{(N)}) (\bi t^{(N-1)},\psi_{M+N-1})} \bigr|^2
     \bigl| c_{(\phi_{N+1},\ldots,\phi_{M+N}) (\bi t^{(N)},\psi_{M+N})} \bigr|^2.
\label{eq:probexpl}
\end{align}
Because there is no summation over the $\bi t$ indices, these diagonal elements are path probabilities of a Markov process. The states correspond to elements of the partition of unity, given by a pair $(\bi\phi,\bi\psi)$ in $I_0$. A pair is determined by $(\phi_1;\psi_1,\ldots\psi_M)$ and, moreover, if $\bi\psi = (1,1,\ldots,1)$ or $\bi\psi = (2,2,\ldots,2)$, then only one $\bi\phi$ can occur. The number of states is thus $2^{M+1}-2$. A transition from state $(\bi\phi,\bi\psi)$ to state $(\bi\phi',\bi\psi')$ is allowed only if the indices $\bi\phi$ and $\bi\psi'$ match, i.e.,
\begin{equation}
\label{eq:match}
 (\phi_2,\ldots,\phi_M) = (\psi'_1,\ldots,\psi'_{M-1}).
\end{equation}
As can be read off from~(\ref{eq:probexpl}), the transition probability from the state $\bigl( (\phi_n,\bi t^{(n)}),\bi\psi \bigr)$ to $\bigl( \bi\phi', (\bi t^{(n)},\psi'_{n+M-1}) \bigr)$ is $\frac{1}{2} \abs{c_{\bi\phi'(\bi t^{(n)},\psi'_{n+M-1})}}^2$ and the initial measure assigns the weight $2^{-M}\, \abs{c_{\bi\phi\bi\psi}}^2$ to $(\bi\phi,\bi\psi)$. Equation~(\ref{eq:punitE}) ensures that these objects indeed correspond to a transition matrix and a probability measure.

We simplify notation and denote the set of states by $A$, the transition probabilities by $P_{ab}$, $a,b\in A$, the initial measure by $\mu$ and the measure after $n$ time steps by $\mu_n$. The entropy we are looking for is,
\begin{align}
 S(\refdens{N})
 &= - \sum_{a_0,a_1,\ldots, a_N}
 \mu(a_0)\, P_{a_0a_1} \cdots P_{a_{N-1}a_N}
 \ln \bigl( \mu(a_0)\, P_{a_0a_1} \cdots P_{a_{N-1}a_N} \bigr)
\nonumber \\
 &= - \sum_{a_0,a_1,\ldots, a_{N-1}}
 \mu(a_0)\, P_{a_0a_1} \cdots P_{a_{N-2}a_{N-1}}
 \ln \bigl( \mu(a_0)\, P_{a_0a_1} \cdots P_{a_{N-2}a_{N-1}} \bigr)
\nonumber \\
 &\phantom{=\ } - \sum_{a_{N-1}a_N} \mu_{N-1}(a_{N-1})\, P_{a_{N-1}a_N}
 \ln P_{a_{N-1}a_N}
\nonumber \\
 &= \cdots
\nonumber \\
 &= - \sum_{n=0}^{N-1} \, \sum_{a\in A} \mu_n(a) \,
 \sum_{b\in A} P_{ab}\ln P_{ab}.
\label{eq:markov1}
\end{align}
We have thus to calculate the quantity $\sum_a \nu(a) \sum_b P_{ab} \ln P_{ab}$ for a measure $\nu$.

{From} a given state $a$ there are only 3 or 4 possible states $b$ to go to. This defines a partition $(A_3,A_4)$ of the set of states $A$. More explicitly, the state $(\bi\phi,\bi\psi)\in A_3$ if and only if $(\phi_2,\ldots,\phi_M)=(111\ldots 1)$ or $(222\ldots 2)$. {From} now on we fix values for the coefficients $c_{\bi\phi\bi\psi}$, namely,
\begin{equation}
\label{eq:cchoice}
 c_{\bi\phi\bi\psi} :=
 \begin{cases}
  1 & \text{if $\bi\psi=(1,1,\ldots,1)$ or $\bi\psi=(2,2,\ldots,2)$} \\
  \frac{1}{\sqrt{2}} &\text{otherwise}.
\end{cases}
\end{equation}
Note that with this choice~(\ref{eq:punitE}) is fulfilled. The transition probabilities from a state $a\in A_3$ are $(\frac{1}{2},\frac{1}{4},\frac{1}{4})$ and from a state $a\in A_4$ $(\frac{1}{4},\frac{1}{4},\frac{1}{4},\frac{1}{4})$. We obtain,
\begin{equation}
\label{eq:markov2}
 - \sum_{a\in A}\nu(a) \, \sum_{b\in A} -P_{ab} \ln P_{ab}
 = \nu(A_3) \, \frac{3}{2}\ln 2 + \nu(A_4) \, 2\ln 2.
\end{equation}

The probabilities $\nu(A_3)$ and $\nu(A_4)$ will be computed by  coarse-graining the set of states $A$. For $p,q\in\{1,2\}$ and $s\in\{0,1,\ldots M\}$, define the sets
\begin{equation*}
 \eset_{pq}^s :=
 \bigl\{ ((p,\bi\phi),(\bi\psi,q)) \,\bigm|\, \bi\phi,\bi\psi\in\{1,2\}^{M-1},\,
 p + \Sigma\bi\phi = \Sigma\bi\psi + q = M + s \bigr\}.
\end{equation*}
We will consider these groups as the states of a new process. Note that there
are only $4M-2$ of them, a number that should becompared with $2^{M+1}-2$ previously. To shorten notation, we shall use $\{C_i\subset A\}$ for the coarse-grained states, and $\{a_i\in A\}$ for the fine-grained states.

No matter how one chooses the map $(\phi_1;\psi_1,\ldots\psi_M) \mapsto  (\phi_2,\ldots,\phi_M)$, these two processes have a peculiar structure. Firstly, for given $a_1$ and $C_2$, if there is a transition possible from $a_1$ to $a_2\in C_2$, then this $a_2$ is unique. Moreover, with the choice~(\ref{eq:cchoice}) for the coefficients $c_{\bi\phi\bi\psi}$, for given $C_1$ and $C_2$, all allowed transitions from $a_1\in C_1$ to $a_2\in C_2$ have the same probability. Therefore, it makes sense to write the transition probabilities as $P_{C_1C_2}$.

Explicitly, these transitions are given as follows: for $s=2,\ldots,M-2$, $(s=1,p=1)$ and $(s=M-1,p=2)$
\begin{equation}
\label{eq:transit1}
\begin{matrix}
 \eset_{11}^s, \eset_{12}^s &\ \to\
 &\eset_{11}^s,\eset_{21}^s,\eset_{12}^{s+1},\eset_{22}^{s+1} \\
 \eset_{21}^s, \eset_{22}^s &\ \to\
 &\eset_{11}^{s-1},\eset_{21}^{s-1},\eset_{12}^s,\eset_{22}^s.
\end{matrix}
\end{equation}
These states have 4 possible transitions, i.e.\ they constitute the set $A_4$. For other combinations $(s,p)$
\begin{equation}
\label{eq:transit2}
\begin{matrix}
 \eset_{11}^0, \eset_{21}^1, \eset_{22}^1&\ \to\
 &\eset_{11}^0, \eset_{12}^1, \eset_{22}^1 \\
 \eset_{22}^M, \eset_{11}^{M-1}, \eset_{21}^{M-1}&\ \to\
 &\eset_{22}^M, \eset_{11}^{M-1}, \eset_{21}^{M-1}.
\end{matrix}
\end{equation}
These states have 3 possible transitions and constitute $A_3$.

The special structure of the considered processes has two important consequences. Firstly, the coarse-grained process is still Markovian. Indeed, the probability for a coarse-grained path is
\begin{align*}
 P(C_1C_2\ldots C_N)
 &= \sum_{a_1\in C_1} \sum_{a_2\in C_2} \cdots \sum_{a_N\in C_N}
 \mu(a_1)\, P_{a_1a_2}P_{a_2a_3} \cdots P_{a_{N-1}a_N} \\
 &= \sum_{a_1\in C_1} \sum_{a_2\in C_2} \cdots \sum_{a_{N-1}\in C_{N-1}}
 \mu(a_1)\, P_{a_1a_2} \cdots P_{a_{N-2}a_{N-1}} P_{C_{N-1}C_N} \\
 &= \cdots \\
 &= \sum_{a_1\in C_1} \mu(a_1)\, P_{C_1C_2}\cdots P_{C_{N-1}C_N} \\
 &= \mu(C_1)\, P_{C_1C_2} \cdots P_{C_{N-1}C_N}
\end{align*}
Secondly, the entropy formula~(\ref{eq:markov2}) for the fine-grained and coarse-grained process leads to the same result. Indeed,
\begin{align*}
 - \sum_a \nu(a) \sum_b P_{ab} \ln P_{ab}
 &= - \sum_{C_1} \sum_{a\in C_1} \nu(a) \sum_{C_2} P_{C_1C_2} \ln P_{C_1C_2} \\
 &= - \sum_{C_1} \nu(C_1) \sum_{C_2} P_{C_1C_2} \ln P_{C_1C_2}.
\end{align*}
In other words, there is no information loss due to the coarse-graining.

Inspecting the transition graph of the coarse-grained process~(\ref{eq:transit1}--\ref{eq:transit2}), one can see that it is strongly connected. Therefore, the transition matrix is irreducible~\cite{HJ}. Moreover, there is a strictly positive diagonal element, which implies that the transition matrix is primitive~\cite{HJ}. As a consequence, it has an unique invariant measure $\mu_\infty$ and the initial measure converges to it. Because we are only interested in the limit $N\to\infty$ in~(\ref{eq:markov1}), we can as well calculate this entropy using the invariant measure from the start. Therefore
\begin{align}
 \s h(\Theta,\omega)
 &\ge \lim_{N\to\infty} \frac{1}{N}\, \s S(\refdens{N})
\nonumber \\
 &= - \sum_{C_1} \mu_\infty(C_1) \sum_{C_2} P_{C_1C_2} \ln P_{C_1C_2}
\nonumber \\
 &= \mu_\infty(A_3) \, \frac{3}{2} \ln 2 + \mu_\infty(A_4) \, 2\ln 2,
\label{eq:markov3}
\end{align}
where we used~(\ref{eq:markov2}) in the last line. Obtaining the values $\mu_\infty(A_3)$ and $\mu_\infty(A_4)$ finishes the computation.

The invariant measure $\mu_\infty$ can be easily calculated and equals
\begin{equation*}
 \mu_\infty(\phi,\psi) =
 \begin{cases}
  \frac{1}{2M} &\text{if } \bi\phi = \bi\psi =(1,1,\ldots,1) \text{ or } \bi\phi =
  \bi\psi = (2,2,\ldots,2) \\
  \frac{1}{4M} &\text{otherwise.}
 \end{cases}
\end{equation*}
Therefore, $\mu_\infty(A_3)=2/M$ and $\mu_\infty(A_4)=(M-2)/M$. Substituting this in~(\ref{eq:markov3})
\begin{equation}
\label{eq:markov4}
 \s h(\Theta,\omega) \ge \frac{2}{M} \, \frac{3}{2} \ln 2 + \frac{M-2}{M} \, 2\ln 2
 = \bigl(2-\frac{1}{M}\bigr) \, \ln 2
\end{equation}
and this converges to $2\ln 2$ when $M$ goes to infinity. We have therefore saturated the upper bound~(\ref{ub2}).

\section{Conclusion}

In this paper, we studied the dependence of the ALF dynamical entropy on the class of allowed partitions. We considered two basic dynamical systems: the shift on a spin chain and the shift on a Fermionic chain. In these cases, the dynamical entropy seems to be robust for natural restrictions on the classes of allowed partitions. These model systems saturate, however, an upper bound following from simple dimensional estimates. It would be interesting to go beyond this situation and investigate examples where this is no longer the case.

\noindent
\textbf{Acknowledgements: }
It is a pleasure to thank R.~Alicki for useful discussions. This work was partially supported by F.W.O. Vlaanderen grant G.0109.01.


\begin{thebibliography}{99}

\bibitem{AF1}
 R.~Alicki and M.~Fannes:
 Defining quantum dynamical entropy,
 \emph{Lett.\ Math.\ Phys.\ }\textbf{32}, 75--82 (1994)

\bibitem{AF2}
 R.~Alicki and M.~Fannes:
 \emph{Quantum Dynamical Systems},
 Oxford University Press, Oxford, 2001

\bibitem{AN}
 R.~Alicki and H.~Narnhofer:
 Comparison of dynamical entropies of non-commutative shifts,
 \emph{Lett.\ Math.\ Phys.\ }\textbf{33}, 241--247 (1995)

\bibitem{CNT}
 A.~Connes, H.~Narnhofer and W.~Thirring:
 Dynamical entropy of C*-algebra and von Neumann algebras,
 \emph{Commun.\ Math.\ Phys.\ }\textbf{112}, 691--719 (1987)

\bibitem{Dav}
 K.R.~Davidson:
 \emph{C*-Algebras by Example},
 American Mathematical Society, Rhode Island, 1996

\bibitem{HJ}
 R.A.~Horn and C.R.~Johnson:
 \emph{Matrix Analysis},
 Cambridge University Press, Cambridge, 1985

\bibitem{FH}
 M.~Fannes and B.~Haegeman:
 Quantum dynamical entropies for classical stochastic systems,
 \emph{Rep.\ Math.\ Phys.\ }\textbf{52}, 151--165 (2003)

\bibitem{P}
 K.~Petersen:
 \emph{Ergodic Theory},
 Cambridge University Press, Cambridge, 1983

\bibitem{ST}
 J.L.~Sauvageot and J.P.~Thouvenot:
 Une nouvelle d\'efinition de l'en\-tro\-pie dynamique des syst\`emes non commutatifs,
 \emph{Commun.\ Math.\ Phys.\ }\textbf{145}, 411--23 (1992)

\end{thebibliography}
\end{document}